\documentclass[letterpaper]{article} % DO NOT CHANGE THIS
\usepackage{aaai23}  % DO NOT CHANGE THIS
\usepackage{times}  % DO NOT CHANGE THIS
\usepackage{helvet}  % DO NOT CHANGE THIS
\usepackage{courier}  % DO NOT CHANGE THIS
\usepackage[hyphens]{url}  % DO NOT CHANGE THIS
\usepackage{graphicx} % DO NOT CHANGE THIS
\usepackage{natbib}  % DO NOT CHANGE THIS AND DO NOT ADD ANY OPTIONS TO IT
\usepackage{caption} % DO NOT CHANGE THIS AND DO NOT ADD ANY OPTIONS TO IT
\usepackage{algorithm}
\usepackage{algorithmic}
\usepackage{amsthm}
\usepackage{amsmath}
\usepackage{multirow}
\usepackage{booktabs}
\newtheorem{definition}{\indent Definition}

\usepackage{graphicx}

\usepackage{url}

\usepackage{xcolor}

\usepackage{newfloat}
\usepackage{listings}
\DeclareCaptionStyle{ruled}{labelfont=normalfont,labelsep=colon,strut=off} % DO NOT CHANGE THIS
\floatstyle{ruled}
\newfloat{listing}{tb}{lst}{}
\floatname{listing}{Listing}
\title{Set-to-Sequence Ranking-Based Concept-Aware Learning Path Recommendation}
\author {
    Xianyu Chen\textsuperscript{\rm 1},
    Jian Shen\textsuperscript{\rm 1},
    Wei Xia\textsuperscript{\rm 2},
    Jiarui Jin\textsuperscript{\rm 1},
    Yakun Song\textsuperscript{\rm 1},
    Weinan Zhang\textsuperscript{\rm 1$\ast$},
    Weiwen Liu\textsuperscript{\rm 2},
    Menghui Zhu\textsuperscript{\rm 1},
    Ruiming Tang\textsuperscript{\rm 2},
    Kai Dong\textsuperscript{\rm 3},
    Dingyin Xia\textsuperscript{\rm 3},
    Yong Yu\textsuperscript{\rm 1}\thanks{Corresponding author.}
}
\affiliations {
    \textsuperscript{\rm 1} Shanghai Jiao Tong University\\
    \textsuperscript{\rm 2} Huawei Noah's Ark Lab\\
    \textsuperscript{\rm 3} Huawei Technologies Co Ltd\\
    \{xianyujun,r\_ocky,jinjiarui97,ereboas,wnzhang,zerozmi7\}@sjtu.edu.cn, yuyong@apex.sjtu.edu.cn\\ \{xiawei24,liuweiwen8,tangruiming,dongkai4,xiadingyin\}@huawei.com
}

\usepackage{bibentry}
\begin{document}
\maketitle

\begin{abstract}
With the development of the online education system, personalized education recommendation has played an essential role.
In this paper, we focus on developing path recommendation systems that aim to generate and recommend an entire learning path to the given user in each session.
% Although a series of existing approaches have been considered ; 
Noticing that existing approaches fail to consider the correlations of concepts in the path,
% Learning path recommendation refers to generating and recommending an entire learning path to the user in one recommendation session, and the process can only be evaluated after the student has completed the entire path. 
% Among a series of existing recommendation algorithms, \emph{path recommendation}
% has become an attractive issue. Among them, learning path recommendation is an important direction.
% The main difficulty in this problem is how to plan a learning path that maximizes the improvement of the target concept within a reasonable time. 
% It is difficult for existing methods to strike a balance between time and effectiveness. 
% To this end, 
we propose a novel framework named \textbf{S}et-to-Sequence \textbf{R}anking-Based \textbf{C}oncept-Aware Learning Path Recommendation (\textbf{SRC}), which formulates the recommendation task under a set-to-sequence paradigm. 
% By modeling the problem as a set-to-sequence problem, SRC can quickly recommend a correct path to students.
Specifically, we first design a concept-aware encoder module that can capture the correlations among the input learning concepts.
The outputs are then fed into a decoder module that sequentially generates a path through an attention mechanism that handles correlations between the learning and target concepts. 
Our recommendation policy is optimized by policy gradient.
% Then, a decoder module based on a sequential network and attention mechanism is used to estimate the benefit of the concept to be learned to the target concept and plan a path accordingly. 
In addition, we also introduce an auxiliary module based on knowledge tracing to enhance the model's stability by evaluating students' learning effects on learning concepts.
We conduct extensive experiments on two real-world public datasets and one industrial dataset, and the experimental results demonstrate the superiority and effectiveness of SRC. Code now is available at  
% \href{https://gitee.com/mindspore/models/tree/master/research/recommend/SRC}
\url{https://gitee.com/mindspore/models/tree/master/research/recommend/SRC}.

\end{abstract}

\section{Introduction}
Different from providing the same learning content for all students in each classroom session in traditional learning,
% (such as a classroom session), 
adaptive learning aims to tailor different learning objectives to meet the individual needs of different learners \citep{carbonell1970ai}. 
Existing recommendation methods of learning content can be summarized into two categories: 
(i) Step by step, the following learning item is recommended for students in real-time, and the interaction of each step (i.e., students' answers) will be integrated into the recommendation for the next step \citep{liu2019exploiting,cai2019learning,huang2019exploring}.
(ii) Plan a certain length of the learning path for students at one time. The latter is because users sometimes want to know the entire learning path at the beginning (for example, universities need to organize courses for students)  \citep{joseph2022exploring,chen2022personalized,shao2021degree,bian2019adaptive,dwivedi2018learning}.
As the latter direction is more restricted and complex (e.g., larger search space, less available feedback), it is more challenging and is also the main focus of this paper.
\begin{figure}[htbp!]
    \centering
    \includegraphics[width=0.48\textwidth]{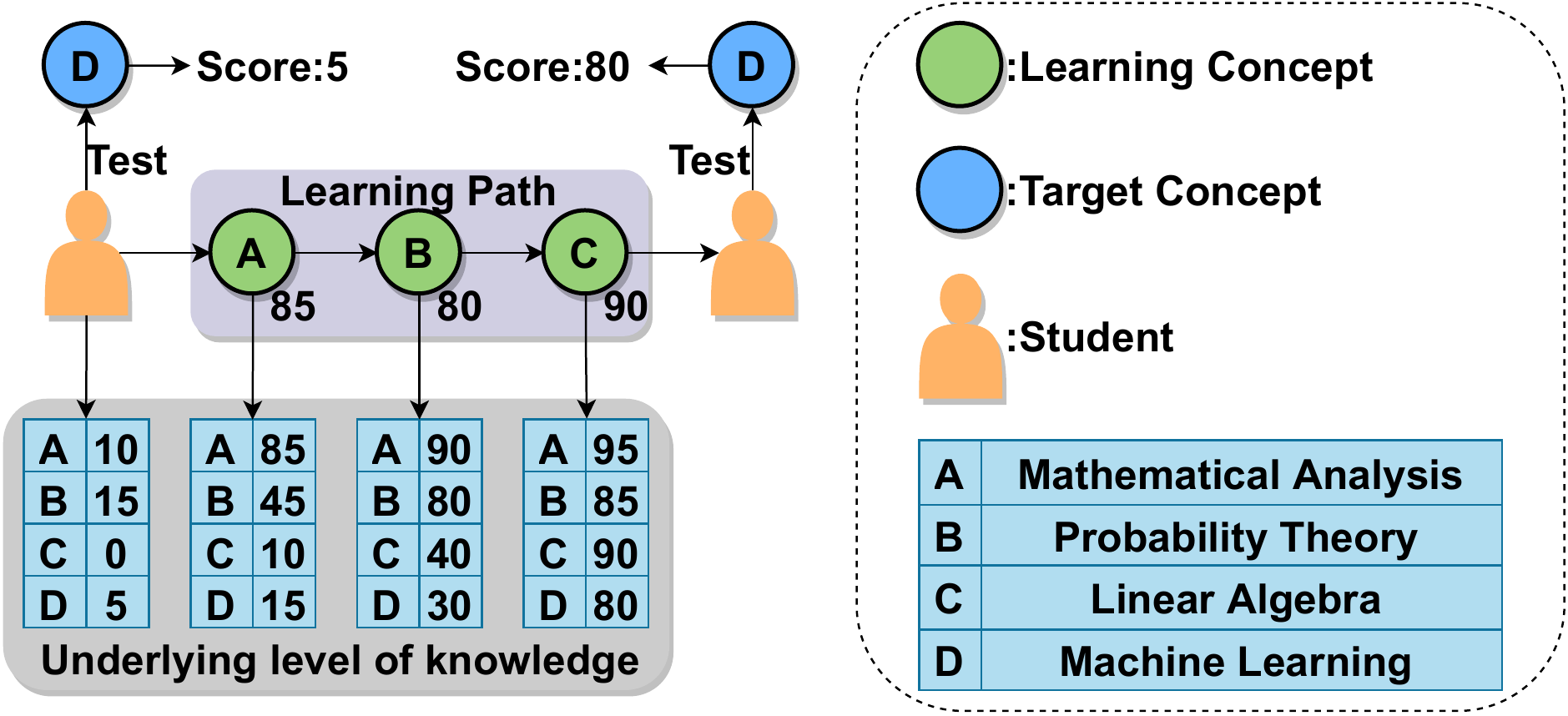}
    \caption{Illustration of the student learning process to improve a student's mastery of concept D by learning the path composed of three concepts A, B, and C.
    The student is given a test before and after the path to knowing his mastery of D. 
    The bottom four tables show the student's potential mastery of all concepts at each step, which is not accessible during the training process.
    % Note that this information cannot be obtained directly. 
    We can only know the mastery of the current learning concept. 
    For example, after learning B, we know his mastery of B is 80.}
    \label{fig:Path}
\end{figure}
Previous studies   \citep{piaget1970genetic,inhelder1976piaget,pinar1995understanding} reveal that cognitive structure greatly influences adaptive learning, which includes both the relationship between items (e.g., premise relationship and synergy relationship) and the characteristics of students' dynamic development with learning. 
Most existing methods to solve learning path planning are either based on a knowledge graph (or some relationship between concepts) to constrain path generation \citep{liu2019exploiting,shi2020learning,wang2022multi}, or based on collaborative filtering of features to search for paths  \citep{joseph2022exploring,chen2022personalized,nabizadeh2019estimating}.
% Although these models have been applied to some extent, 
However, these models can not penetrate into the important features of the cognitive structure perfectly, and the model is relatively simple, resulting in the path generated either with a low degree of individuation or with a poor learning effect.
From this perspective, we argue that \emph{how to effectively mine the correlations among concepts and the important characteristics of students in learning path planning} is still challenging, and summarize the specific challenges as follows:

\begin{itemize}
\item \textbf{(C1)} How to effectively explore the correlations among concepts?
There may be complex and diverse correlations between concepts, such as prerequisite relationship and synergy relationship, which will affect students' learning of concepts \citep{tong2020structure,liu2019exploiting}. As shown in Figure~\ref{fig:Path}, mastery, of course, A (Mathematical Analysis) is of greater help to mastery of course B (Probability Theory), and of less help to mastery, of course, C (Linear Algebra). Therefore, it should be taken into account when planning the learning path.
\item \textbf{(C2)}
How to evaluate and optimize the generation algorithm by effectively using the students' learning effect on the target concepts? As shown in Figure\ref{fig:Path}, we expect students to achieve the best improvement in the target concept D (Machine Learning). 
However, the existing path recommendation algorithms either do not use this feedback but use indirect factors such as similarity degree and occurrence probability \citep{joseph2022exploring,shao2021degree}, or lack of excellent generation algorithms \cite{zhou2018personalized,nabizadeh2019estimating}. As a result, it is difficult for them to provide an efficient learning path. 
This is because it is still challenging to optimize a path using only feedback that is available at the end of the path. 
In contrast, in the stepwise recommendation scenario, immediate feedback can be obtained at the end of each step, which allows some more advanced reinforcement learning (RL) algorithms \citep{sun2021cost,li2021optimal} to be applied.
\item \textbf{(C3)}
How can student feedback on learning concepts be incorporated into the model? As shown in Figure \ref{fig:Path}, students have different learning feedback for concepts A, B, and C on the path after learning. In the field of knowledge tracing (KT), this information plays a great role in modeling students' knowledge levels. Many models \citep{piech2015deep,yang2020gikt,zhang2017dynamic} take students' past answers as features to predict the current answer.
In  \citep{liu2019exploiting}, its DKT  \citep{piech2015deep} module used this information to trace students' knowledge levels in real-time to adjust recommendations for the next step. However, in path recommendation, this feedback can only be obtained after the path ends, so the above approach is difficult to implement here.
\end{itemize}

To address these challenges, we propose a novel framework Set-to-Sequence Ranking-based Concept-aware Learning Path Recommendation (\textbf{SRC}). We formulate the learning path recommendation task as a set-to-sequence paradigm.
In particular, firstly, in order to mine the correlation between concepts (\textbf{C1}), we design a concept-aware encoder module. This module can globally calculate the correlation between each learning concept and other learning concepts in the set so as to get a richer representation of the concept.
At the same time, in the decoder module, on the one hand, we use a recurrent neural network to update the state of students; on the other hand, we use the attention mechanism to calculate the correlation between the remaining learning concepts in the set and the target concepts, so as to select the most suitable concept in the current position of the path.
Secondly, we need to effectively utilize feedback on the target concepts (\textbf{C2}). Since the feedback is generally continuous and considering the large path space, the policy gradient algorithm is more suitable in this case. Thus the correlation between the learning concept and the target concepts calculated by the previous decoder can be expressed in the form of selection probability. So we get a parameterized policy, and we can update the model parameters in a way that maximizes the reward.
Finally, we designed an auxiliary module to utilize feedback on learning concepts (\textbf{C3}). Similar to the KT task, the student state updated by the previous decoder at each step is fed into an MLP to predict the student's answer at that step. In this way, students' feedback on the learning concepts can participate in the updating of model parameters to enhance the stability of the algorithm.
\section{Related Works}
\textbf{Learning Path Recommendation.} A class of branches \citep{joseph2022exploring,chen2022personalized,zhou2018personalized, nabizadeh2019estimating,liu2020learning,shao2021degree} in existing methods models the task as a general sequence recommendation task, which is dedicated to reconstructing the user's behavior sequence. For example, in \citet{zhou2018personalized}, KNN \citep{cover1967nearest} is used to complete the collaborative filtering and then RNN is used to estimate the learning effect; in \citet{shao2021degree}, the BERT \citep{devlin2018bert} paradigm is directly used to solve this problem.
Another branch \citep{liu2019exploiting,shi2020learning,wang2022multi,zhu2018multi} focuses on mining the role of knowledge structure. For example, \citet{zhu2018multi} formulates some rules based on the knowledge structure to constrain the generation of paths. In general, most of the above methods fail to take full advantage of student feedback on the target concepts. One of the better ones is \citet{liu2019exploiting}, which uses this feedback through reinforcement learning methods to optimize the generative model. However, on the one hand, it can obtain real-time feedback on learning concepts, and its application scenario is actually a step-by-step recommendation; on the other hand, it uses the concept relationship graph as a rule to constrain path generation without mining deeply into their correlations, which makes it challenging to apply in the general case. 
In our method, we use an attention mechanism to mine inter-concept correlations and make full use of various feedback from students to optimize the modeling of correlations, which makes our method more general.

\textbf{Learning Item Recommendation.}
In step-by-step learning item recommendations, immediate feedback is available. This allows them \citep{cai2019learning,huang2019exploring,sun2021cost,li2021optimal} to use more complex RL algorithms. Such as \citet{sun2021cost} use the DQN\citep{mnih2013playing}, and \citet{cai2019learning} use the Advantage Actor-Critic.
Our method also uses policy gradient in RL for optimization, but since we have no immediate feedback, only delayed feedback after the path ends, training may be more difficult. Therefore we introduce the KT auxiliary task to enhance the model stability.

\textbf{Set-to-Sequence Formulation.}
The set-to-sequence task aims to permute and organize a set of unordered candidate items into a sequence whose solutions can be roughly divided into three fields: point-wise, pair-wise, and list-wise. 
Among them, the point-wise method is the most widely used, which is designed to score each item individually, and then rank the items in descending order of their scores \citep{friedman2001greedy}. 
The pair-wise methods  \citep{burges2005learning,joachims2006training} do not care about the specific score of each item. Instead, formulate the problem pair-wise, focusing on predicting the relative orders among each pair of items. 
% that compares any two items with a higher degree of correlation. 
The list-wise algorithms \citep{burges2010ranknet,cao2007learning,xia2008listwise} treat the entire sequence as a whole, which allows the model to mine the deep correlations among the items carefully. 
Noticing that students' feedback on a concept is likely to be significantly affected by the other concepts on the same path, we here design our model in a list-wise manner.

% And as we discussed earlier, there are complex correlations between concepts, which makes it difficult to evaluate a concept's score independently of other concepts in the path; on the other hand, students' feedback on the target concepts is for the whole paths, and it is difficult to conclude that one concept is more relevant than another. Based on these considerations, the list-wise method is more suitable.

The main difficulty with list-wise is that the sorting process is not completely differentiable because there are no gradients available for sorting operations \citep{xia2008listwise}. One solution randomly optimizing the ranking network by continuous relations \citep{grover2019stochastic,swezey2020pirank}. Another class of branches, named Plackett-Luce (PL) ordering model  \citep{burges2010ranknet,luce2012individual,plackett1975analysis}, represents ordering as a series of decision problems, where each decision is made by softmax operation. Its probabilistic nature leads to more robust performance \citep{bruch2020stochastic}, but computing the gradient of the PL model requires iterating over every possible permutation. A solution proposed in the recent literature \citep{oosterhuis2021computationally} is the policy gradient algorithm   \citep{williams1992simple}.
\section{Problem Formulation}
Consider a student $u$, whose historical sequence of concepts learning is $H=\{h_1,h_2,\cdots,h_k\}$. The record $h_t=\{c_t,y_t\}$ of each time $t$ includes the learned concept $c_t$ and the degree of mastery $y_t$ of the concept. 
Now given a set $S=\{s_1,s_2,\cdots,s_m\}$ consisting of $m$ candidate concepts, the student $u$ is to learn $n$ non-repetitive concepts from $S$ in some order (hence $m\geq n$). Through the study of such a learning path $\pi=\{\pi_1,\pi_2\cdots, \pi_n\}$, he can improve his mastery of some target concepts $T=\{t_1,t_2,\cdots \}$. Following \citep{liu2019exploiting}, we quantify the learning effect as
\begin{equation}
\label{eqn:et}
E_T=\frac{E_e-E_b}{E_{sup}-E_b}
\end{equation}
where $E_e$ and $E_b$ represent the student's mastery of the target concepts before and after the path $\pi$ (which can be obtained through exams), and $E_{sup}$ represents the upper bound of mastery. At the same time, we can also observe the students' mastery $Y_\pi=\{y_{\pi_1},y_{\pi_2},\cdots,y_{\pi_n}\}$ of learning concepts after the end of the path. Then, we can 
formulate our problem as follows:
\begin{definition}
\textbf{Learning Path Recommendation.} Given a student's historical learning sequence $H$, target concepts $T$, and candidate concepts set $S$, it is required to select $n$ concepts from $S$ without repetition and rank them to generate a path $\pi$ to recommend to the student. The ultimate goal is to maximize the learning effect $E_T$ at the end of the path.
\end{definition}
\section{SRC Framework}
\begin{figure*}[htbp!]
    \centering
    \includegraphics[width=\textwidth]{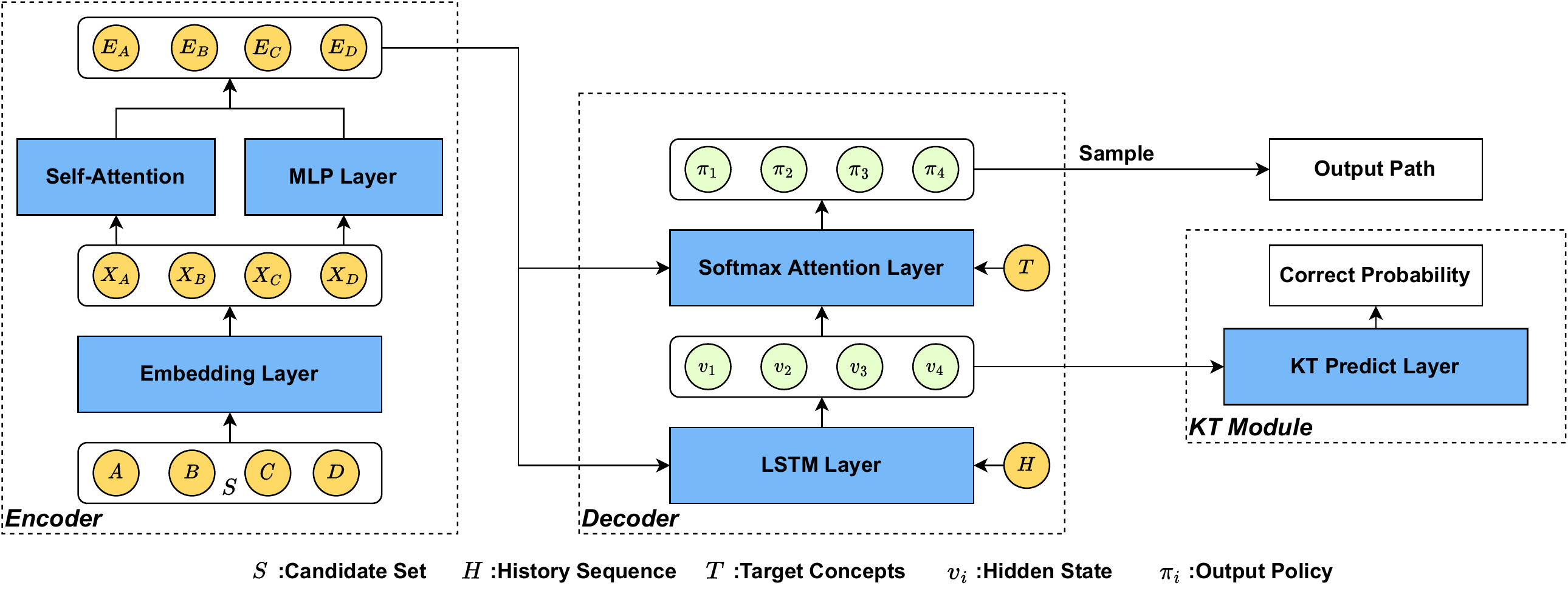}
    \vspace{-7mm}
    % \captionsetup{font=large}
    \caption{The overview of our framework. SRC is composed of the encoder, decoder, and KT auxiliary module. The encoder captures the correlation between concepts in the candidate set $S$ to obtain the representation of concepts $E_S$. The decoder generates a ranking of S based on the information of $E_S$, $T$, and $H$, and outputs the policy $\pi$. KT auxiliary module is responsible for predicting the correct probability of each step on the path.}
    \label{fig:framework}
\end{figure*}
Figure \ref{fig:framework} shows the overview framework of our SRC model. As shown, first we design a concept-aware encoder to model the correlations among candidates' learned concepts to obtain their global representations. Then in the decoder module, we use the recurrent neural network to model the knowledge state of the students along the path and calculate the correlation between the learning concepts and the target concepts through the attention mechanism to determine the most suitable concept for the position. In addition to this, based on the knowledge state obtained in the decoder, we further predict the student's answer to the learning concepts. At the end of the learning path, we pass the obtained feedback $E_T$ and $Y_\pi$ to the model to optimize the parameters.

\subsection{Encoder}
First, for each concept $s_i$ in the candidate concept set $S$, we access the embedding layer to obtain its continuous representation \textbf{$x_{s_i}$}. 
However, as we discussed in the introduction, there are complex and diverse correlations between concepts, and these correlations can seriously affect the final learning effect of the path. 
However, the embedding representation we obtained can only reflect the characteristics of the concept itself in isolation, and cannot reflect the correlation between concepts. Therefore, we need a function $f^e$ to capture these correlations within the set and fuse them into the concept representation to get the global representation $E_s$:
\begin{equation}
    E_s=f^e(X_s), X_s=[x_{s_1},x_{s_2},\cdots, x_{s_m}]^T.
    \label{eqn:encoder}
\end{equation}
For the implementation of $f^e$, a simple approach is to add these concept representations to each concept after a pooling operation (e.g., average pooling operation), unfortunately, it is not capable to model complex correlations. 
Recent literature \citep{pang2020setrank,lee2019set} uses a more complex Transformer to extract information, but it mainly focuses on correlation and thus pays less attention to the unique characteristics of the concept itself. 
And note that our training is based on the paradigm of a policy gradient with only one reward per path. 
With label sparsity, complex models like Transformer are extremely difficult to train due to potential over-smoothing issues \citep{liu2021trap}.
We empirically verify it in our follow-up experiments, which motivates us to combine the above two approaches. 

First, we apply the self-attention mechanism to $X_S$: 
\begin{equation}
    E^a_s=softmax(\frac{QK^T}{\sqrt{d}})V,
    \label{eqn:attention}
\end{equation}
where \begin{equation}
    Q=X_sW^Q, K=X_sW^K, V=X_sW^V, 
\end{equation}
$W^Q,W^K,W^V$ are all trainable weights, $d$ is the dimension of the embedding. At the same time, we pass a simple multi-layer perceptron(MLP) to the embedding and add the average pooling part:
\begin{equation}
    E^l_s=E^l+\frac{\sum_i^me^l_i}{m}, E^l=f^l(X_S),
    \label{eqn:mlp}
\end{equation}
where $f^l$ is the MLP, $e^l_i$ is the feature of the $i$-th concept in $E^l$. Then the final representation of learning concepts in $S$ is 
\begin{equation}
    E_s=[E_s^a;E_s^l],
    \label{eqn:concat}
\end{equation}
where $[.;.]$ is the concat operation. In this way, we obtain the representation $E_s$ that are being aware of the other concepts in the set and retain their own characteristics.

\subsection{Decoder}
After obtaining a representation of each learned concept, we will generate their permutation and the probability in the decoder module:
\begin{equation}
    \pi , P =f^d(E_s, H, T).
    \label{eqn:decoder}
\end{equation}
The implementation of $f^d$ refers to Pointer Network \citep{vinyals2015pointer}. 
First, we design an LSTM (denoted as $g$) \citep{hochreiter1997long} to trace student states. The initial state $v_0$ of the student in $g$ before the start of the path should be related to the student's past learning sequence $H$. Considering that each step $i$ in $H$ contains both the learning concept $c_i$ and the mastery degree $y_i$, $v_0$ is calculated as:
\begin{equation}
    v_o=g([x_{c_1};y_1]W_h,[x_{c_2};y_2]W_h,\cdots, [x_{c_k};y_k]W_h),
    \label{eqn:initial}
\end{equation}
where $x_{c_i}$ is the embedding of concept $c_i$, $W_h$ is a trainable matrix that transforms $[x_{c_1};y_1]$ into the same input dimension as $E_s$. 

Now we assume that the state after the $(i-1)$-th concept $\pi_{i-1}$ in the learning path is $v_{i-1}$. Let $\pi_{<i}$ denote the set of learned concepts, namely $\pi_{<i}=\{\pi_1,\pi_2,\cdots,\pi_{i-1}\}$. Then we calculate the probability distribution of step $i$ as follows:
\begin{equation}
\label{eqn:score}
    d_i^j=w_1^T(W_1v_{i-1}+W_2e_s^j+W_3x_T+b),
\end{equation}
\begin{equation}
\label{eqn:pro}
    P(\hat{\pi_i}=s_j)=\left \{
    \begin{aligned}
        &e^{d_i^j} /
        \sum_{s_k \in S/\pi_{<i}}e^{d_i^k}, 
        \text{ if } s_j \not\in \pi_{<i}\\
	&0, \text{ otherwise}.
    \end{aligned}
    \right.,
\end{equation}
where $j=1,2\cdots,m$; $s_j$ represents the $j$-th concept in $S$ and $e_s^j$ represents its representation obtained by encoder; $x_T$ represents the fusion of embeddings of concepts in T; $w_1,W_1,W_2$ and $b$ are learnable weights or matrices. 
As shown in Eq.~(\ref{eqn:score}), we comprehensively consider student knowledge states, learning concepts, and target concepts to calculate the score of each learning concept under the current step. 
Then in Eq.~(\ref{eqn:pro}), we further use softmax to calculate the probability among the remaining concepts, and the selected concepts are set to 0. Then according to the obtained probability distribution$P(\hat{\pi_i})$, we can sample it to get the concept $\pi_i$ of position $i$. And then we can update the state $v_i$ accordingly:
\begin{equation}
    v_i=g(e_s^{\pi_i},v_{i-1}).
    \label{eqn:update}
\end{equation}

According to the above method, we generate the final path $\pi=\{\pi_1,\cdots,\pi_n\}$ step by step, and the probability $P=\{p_{\pi_1},\cdots,p_{\pi_n}\}$ corresponding to each step. This path will be recommended to students later.

\subsection{Knowledge Tracing Auxiliary Module}

In the decoder, we use LSTM to trace the student state to evaluate the current best-fit learning concept. However, unlike the general practice \citep{piech2015deep,liu2019exploiting}, we have no way to get instant access to the mastery of the student's previous step. Lack of utilization of this feedback may affect the performance of the decoder. 
To this end, we developed this module to predict mastery in a student's process, which acts as an auxiliary task to enhance the reliability and stability of other modules. Specifically, for the $i$-th concept $\pi_i$ on the path, we predict that the probability $p^k_{\pi_i}$of successful mastery by students is:
\begin{equation}
    p^y_{\pi_i}=Sigmoid(f^y(v_i)),
    \label{eqn:KT}
\end{equation}
where $f^y$ denotes a MLP, $Sigmoid()$ denotes the $sigmoid$ function.
\subsection{Optimization Objective}
We build a student simulation to obtain $E_e$, $E_b$, and $E_{sup}$, which will be further specified in the experiment section.
Then, we are able to compute $E_{T}$ according to Eq.~(\ref{eqn:et}).
We treat it as the reward in RL and formulate the loss of 
% part of the loss based on the learning effect ETE_{T} of the target concepts.
% then the loss of 
policy gradient for the path as:
\begin{equation}
\begin{aligned}
    L_\theta &= -E_T\log\prod_i^n p_{\pi_i}
             = -E_T\sum_i^n\log p_{\pi_i}.
\end{aligned}
\end{equation}
Besides $L_\theta$, we also introduce a cross-entropy between the predicted probability $p^y_{\pi_i}$ in the knowledge tracing auxiliary module and the actual feedback $y_{\pi_i}$ of learning concept:
\begin{equation}
    L_y = -\sum_i^n (y_{\pi_i}\log p^y_{\pi_i}+(1-y_{\pi_i})\log(1-p^y_{\pi_i})).
\end{equation}
By combining the above two losses, we can obtain the final loss of the full path:
\begin{equation}
    L=L_\theta +\beta L_y,
    \label{eqn:loss}
\end{equation}
where $\beta$ can be 0 or 1, which is used to control whether the KT task is used to assist the training. 
Algorithm~\ref{algorithm:SRC} shows the training procedure of our method.

\begin{algorithm}[tbhp!]
	\caption{SRC} 
	\label{algorithm:SRC} 
	\begin{algorithmic}[1]
		\STATE Randomly initialize the learning parameters.
		\WHILE{not converged}
		\STATE Randomly sample $T,H,S$.
		\STATE Calculate representation $E_s$ of concepts in $S$ (Eq.~(\ref{eqn:encoder})).
		\STATE Generate path $\pi$ and probability $P$ (Eq.~(\ref{eqn:decoder})).
		\STATE Predict feedback $P_\pi^y$ on the learning path (Eq.~(\ref{eqn:KT})).
		\STATE Get feedback $E_T,Y_\pi$ from student after learning $\pi$. 
		\STATE Compute the gradient and update the parameters w.r.t the loss $L$ (Eq.~(\ref{eqn:loss})).
		\ENDWHILE 
	\end{algorithmic} 
\end{algorithm}
\section{Experiment}
In this section, we detail our experimental setup and results. We also do some discussions and extended investigations to illustrate the effectiveness of our model.

\subsection{Dataset}
Our experiments are performed on two real-world public datasets: ASSIST09\footnote{https://sites.google.com/site/assistmentsdata/home/2009-2010-assistment-data} \citep{feng2009addressing} and Junyi\footnote{https://www.kaggle.com/datasets/junyiacademy/learning-activity-public-dataset-by-junyi-academy} \citep{chang2015modeling}. Some statistics of the two datasets are shown in Table~\ref{table:Dataset}.

\begin{table}[htbp!]
\centering
\resizebox{0.8\linewidth}{!}{
\begin{tabular}{lccc} 
\toprule
Dataset  & \#students & \#concepts & \#average length \\ \midrule
ASSIST09 & 3841       & 167        & 49.54             \\ %\midrule
Junyi    & 5002       & 712        & 54.19             \\ \bottomrule
\end{tabular}}
\caption{Dataset}
\label{table:Dataset}
\end{table}

However, there are some problems with using these two datasets directly to evaluate the model. Specifically, their data are all static, i.e. only answers to concepts that have been answered by students beforehand are known. Our model and other baseline models need to generate new paths and learn student feedback on them for training and evaluation. Therefore, the static original dataset cannot meet our requirements.

\begin{table*}[htbp]
\centering
\resizebox{0.8\textwidth}{!}{
\begin{tabular}{@{\extracolsep{2pt}}ccccccccccc}
    \toprule
    & \multirow{2}{*}{Model} & \multicolumn{4}{c}{ASSIST09} & \multicolumn{4}{c}{Junyi} \\ 
    
    \cmidrule{3-6}
	\cmidrule{7-10}
    & {} & p=0 & p=1 & p=2 & p=3 & 
    p=0 & p=1 & p=2 & p=3 \\\midrule
    \multirow{6}{*}{DKT} & Random  & 
    0.0707 & 0.0995 & 0.1159 & 0.1290 & 
    -0.0854 & -0.0054 & -0.0091 & -0.0075 \\ \cmidrule{2-10}
    & Rule-based & 
    \underline{0.1675} & 
    \underline{0.2070} & 
    \underline{0.1950} & 
    \underline{0.4233} &  
    0.0525 & 
    \underline{0.0969} & 
    \underline{0.1115} & 
    \underline{0.3481}  \\ \cmidrule{2-10}
    & MPC & 
    0.0834 & 0.1163 & 0.1293 & 0.1399 & 
    -0.0584 & 0.0176 & 0.0121 & 0.0209 \\ \cmidrule{2-10}
    & GRU4Rec & 
    0.0862 & 0.1505 & 0.1682 & 0.0755 &  
    -0.0390 & 0.0112 & 0.0152 & -0.1394 \\ \cmidrule{2-10}
    & DQN & 
    0.1215 & 0.1767 & 0.0723 & 0.2949 & 
    \underline{0.0713} & 0.0234 & -0.0118 & 0.2023 \\ \cmidrule{2-10}
    & \textbf{SRC} & 
    \textbf{0.3135}$^*$ & 
    \textbf{0.2971}$^*$ & 
    \textbf{0.2345}$^*$ & 
    \textbf{0.5567}$^*$ & 
    \textbf{0.2555}$^*$ &  
    \textbf{0.1761}$^*$ &
    \textbf{0.1508}$^*$ & 
    \textbf{0.5809}$^*$\\ \midrule
    
    \multirow{6}{*}{CoKT} & Random  & 
    0.0858 & 0.0932 & 0.0917 & 0.0968 & 
    -0.1022 & -0.0700 & -0.0664 & -0.0773 \\ \cmidrule{2-10}
    & Rule-based & 
    0.0928 & 0.1010 & 0.0960 & 0.0990 &  
    -0.0988 & -0.0580 & \underline{-0.0503} & -0.0522  \\ \cmidrule{2-10}
    & MPC & 
    0.1145 & 0.1056 & 0.0918 & 0.1035 & 
    -0.0568 & -0.0699 & -0.0823 & -0.0576 \\ \cmidrule{2-10}
    & GRU4Rec & 
    0.1334 & 0.1242 & \underline{0.1240} & 0.0589 &  
    -0.0333 & -0.0675 & -0.0659 & -0.1385 \\ \cmidrule{2-10}
    & DQN & 
    \underline{0.1403} & 
    \underline{0.1281} & 0.0710 & \underline{0.1253} & 
    \underline{-0.0145} & \underline{-0.0524} & -0.0691 & \underline{0.0781} \\ \cmidrule{2-10}
    & \textbf{SRC} & 
    \textbf{0.1885}$^*$ & 
    \textbf{0.1559}$^*$ & 
    \textbf{0.1574}$^*$ & 
    \textbf{0.2340}$^*$ & 
    \textbf{0.0569}$^*$ &  
    \textbf{-0.0238}$^*$ &
    \textbf{-0.0360}$^*$ & 
    \textbf{0.1709}$^*$\\ \bottomrule
\end{tabular}
}
\captionsetup{font=normalsize}
\caption{Performance comparison of different models under four scenarios for two simulators built on each dataset. * indicates p\textless0.001 in significance tests compared to the best baseline.}
\label{tab:overall}
\end{table*}
To this end, we refer to some of the practices of \citep{liu2019exploiting,hu2018reinforcement}, and design a simulator that can dynamically assess students' knowledge level and return feedback. Specifically, this simulator is data-driven. We train the KT model on static data. The input of the model is the student's past learning sequence, and the output is the current concept answer probability. After the simulator training is completed, we can use it to simulate the learning situation of students on the paths recommended by various models and obtain the corresponding $E_T$ to complete the effect evaluation. To enhance the reliability of the experiments, we use two KT models: DKT \citep{piech2015deep} and CoKT \citep{long2022improving} to build different simulators.

Meanwhile, to be able to compare the performance of various models more comprehensively, we formulate 4 different sources of candidate concept set $S$. Specifically, if the recommended path length is $n$, then the source of $S$ will be: \textbf{1.} A fixed number of $n$ concepts; \textbf{2.} Group all concepts, each with a size of $n$, and randomly select one group at a time; \textbf{3.} Randomly select $n$ concepts each time; \textbf{4.} All concepts. Of course, the division in the first two sources is consistent for all models. In the experimental results reported later, $p=0, 1, 2, 3$ represents these four sources in turn.

\subsection{Baselines}
To demonstrate the effectiveness and robustness of our framework, we compare it with the following methods:
\begin{itemize}
    \item \textbf{Random}: Randomly select concepts from $S$ and then randomly arrange the generated paths.
    \item \textbf{Rule-based}: Let the simulator return the learning effect of the target concepts after learning each concept separately in $S$. Then, according to this effect, the concepts are sorted from smallest to largest to generate paths.
    \item \textbf{MPC}: Using the Model Predictive Control \citep{deisenroth2011pilco} in RL combined with KT, each step predicts the effect of several random search paths and makes the current action.
    \item \textbf{GRU4Rec}: A classic sequential recommendation model \citep{hidasi2015session}. After inputting the past sequence, predict the probability distribution of the next concept. Note that it is learned based on the original dataset.
    \item \textbf{DQN}: Classic reinforcement learning model \citep{mnih2013playing}. Here we pre-train a DKT model based on the original data to generate the required state in DQN.
\end{itemize}
\subsection{Experiment Setting}
The learning rate is decreased from the $1\times10^{-3}$ to $1\times10^{-5}$ during the training process. The batch size is set as $128$. The weight for L2 regularization term is $4\times10^{-5}$. The dropout rate is set as $0.5$.
%Dropout is mainly used for the MLP of the final output probability in KT, where 0.5 is an acceptable value
The dimension of embedding vectors is set as $64$. All the models are trained under the same hardware settings with 16-Core AMD Ryzen 9 5950X (2.194GHZ), 62.78GB RAM, and NVIDIA GeForce RTX 3080 cards.
% All code will be released with \citep{mindspore}. 

\subsection{Experiment Result}
Table~\ref{tab:overall} shows the performance comparison between SRC and other baseline models in their respective cases, where the evaluation index is the learning effect $E_T$, and the path length is $20$. From these results, we have the following findings.
\begin{itemize}
    \item Our model SRC outperforms all baselines in any case. This may be because our model adequately models inter-concept correlations and effectively utilizes feedback.
    \item In DKT, the rule-based method generally achieves the best performance compared to other baselines. 
    However, in CoKT, the performance of this method is poor, and most cases are close to the random method. Moreover, under CoKT, in many cases even SRC its $E_T$ is negative and the maximum value of $E_T$ learned by various models is also lower than the value under DKT.
    This may reflect the difference in the properties of the two simulators, such as the consideration of the relationship between concepts, the speed of forgetting knowledge, etc. Overall, learning under CoKT is more difficult.
    \item GRU4Rec performs well in some cases. This shows that the original learning sequences of students have some value and can reflect the relationship between concepts to some extent. Except that the performance is even worse than random in the case of $p=3$, which may be because this kind of scenario of choosing the optimal path from all concepts requires a wider range of collocations, and the limited raw data cannot extract this paradigm.
    \item DQN performs well in most cases (3rd in DKT, 2nd in CoKT). This reflects the superiority of interaction-based reinforcement learning methods in this setting. Its poor performance at $p=2$ may be because the situation is more complex and the exploration of DQN is insufficient.
\end{itemize}

\subsection{Ablation Study}

\textbf{Impact of Encoder and KT Auxiliary Module.} Table~\ref{tab:Module} 
shows the performance comparison of various variants of SRC, where SRC$_A$ and SRC$_M$ represent the case where the encoder of SRC only uses self-attention or MLP, respectively. 
SRC$^-$ means $\beta=0$ during training, i.e. no KT auxiliary module is used. First of all, it can be seen that the performance of the model is degraded after replacing the original encoder with or without the KT module. This shows that the encoder in SRC combining self-attention and MLP indeed retains the advantages of both, which not only mines the correlation between concepts but also retains its own features.

Then, note that in SRC$_A^-$, after removing the KT module, a very large drop in model performance occurs compared to SRC$_A$, far exceeding that in the other two cases. In addition to this, SRC$_A^-$ sometimes does not converge in experiments that other models have never experienced. This validates our previous concern that such complex networks would be significantly more difficult to train under this sparse reward reinforcement learning paradigm. Therefore, after using the KT module, the student feedback in $Y_\pi$ not only brings more information, but also reduces the training difficulty of the encoder, and the performance is greatly improved. In SRC and SRC$_M$, there is no encoder training problem, so the performance improvement is not so obvious.

\begin{table}[htbp!]
    \centering
    \begin{tabular}{@{}lcccc@{}}
    \toprule
    Model & p=0 & p=1 & p=2 & p=3 \\
    \midrule
        SRC$_M^-$ &  0.2881 & 0.2878 &  0.2276 & 0.5391 \\ 
        SRC$_M$ &  0.3098 & 0.2954 &  0.2311 & 0.5405 \\ 
        \midrule
        SRC$_A^-$ &  0.2234 & 0.2018 &  0.1587 & 0.4539 \\ 
        SRC$_A$ &  0.2943 & 0.2891 &  0.2301 & 0.5378 \\ 
        \midrule
        SRC$^-$ &  0.3025 & 0.2910 &  0.2291 & 0.5290 \\ 
        SRC &  
        \textbf{0.3135}$^*$ & 
        \textbf{0.2971}$^*$ & 
        \textbf{0.2345}$^*$ & 
        \textbf{0.5567}$^*$   \\ 
    \bottomrule
    \end{tabular}
    \caption{Performance comparison of different variants of SRC under DKT built on ASSIST09.}
    \label{tab:Module}
\end{table}

\begin{figure}[htbp!]
\centering
\includegraphics[width=0.48\textwidth]{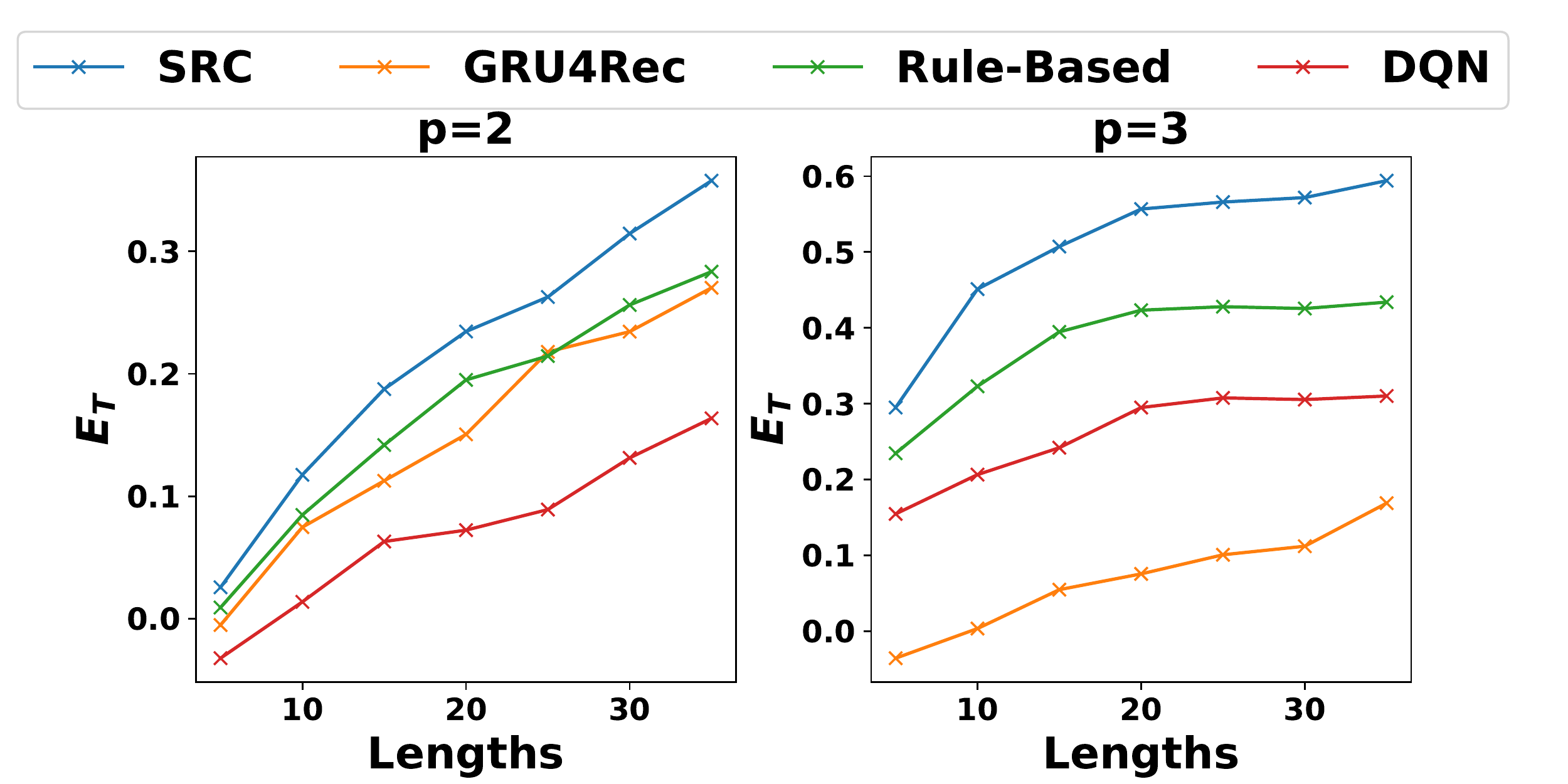}
% \begin{minipage}[t]{0.32\textwidth}
% \centering
% \includegraphics[width=\textwidth]{fig/length_1.pdf}
% \end{minipage}
% \begin{minipage}[t]{0.32\textwidth}
% \centering
% \includegraphics[width=\textwidth]{fig/length_2.pdf}
% \end{minipage}
\caption{Impact of Length}
\label{fig:Length}
\end{figure}

\textbf{Effects of different path lengths.}
Figure~\ref{fig:Length} shows the performance of paths of various lengths generated by different models under two complex scenarios of $p = 2,3$. 
First, the performance rankings of various models are basically unchanged at various lengths, further illustrating the effectiveness of our model. 
Then, the learning effect $E_T$ all grows with the length of the path, which is also in line with the intuition in education. Also note that in the $p=3$ scenario, the performance growth of all models becomes very slow after length$>20$. 
This is probably because the concepts that make up the path in this scenario are selectable by the model. While the number of concepts that are helpful for learning the target concept is limited, they are already selected when the path is short. Concepts added on longer paths have little value and are offset by factors such as forgetting.

\subsection{Result for Industrial Dataset}
To verify the effectiveness of SRC in practical applications, we deploy our model in the online education department of Huawei. Table~\ref{table:Industrial} shows some experimental results on the company's internal industrial dataset. The dataset includes 159 students and 614 concepts and the average length of trajectories is 108.99. 

It can be seen that SRC shows the best performance. The performance of other methods is also similar to those on public datasets. The main difference is that the negative rewards appear more frequently here. Under DKT, the random method even has negative rewards close to $-1$; correspondingly, under DKT, the SRC method can learn rewards close to the upper bound of $1$. Under CoKT, although the reward of the learned optimal path is still negative in some cases, the fluctuation range is greatly reduced. These results reflect the different properties of difficulty curves and forgetting curves under different simulators built on different datasets. And our model can show the best performance in a variety of situations, indicating the effectiveness and generalization of our method.

Further online experimental results are being deployed and collected in the coming weeks.
\begin{table}[htbp!]
    \centering
    \resizebox{0.98\linewidth}{!}{
    \begin{tabular}{@{}lcccccc@{}}
    \toprule
         & Model & p=0 & p=1 & p=2 & p=3  \\
    \midrule
        \multirow{4}{*}{DKT} & Rule-based & 0.0319  & 0.2092  & 0.1622  & 0.9507   \\ 
        \cmidrule{2-6} & Random & -0.8202  & -0.7088  & -0.8098  & -0.8885  \\
        \cmidrule{2-6} & DQN & 0.4495  & -0.2504  & -0.6021  & 0.8800   \\ 
        \cmidrule{2-6} & SRC & \textbf{0.9319*}  & \textbf{0.4701*}  & \textbf{0.2842*}  & \textbf{0.9861*}  \\ 
    \midrule
        \multirow{4}{*}{CoKT} & Rule-based & -0.0445  & -0.0631  & -0.0819  & -0.0595   \\ 
        \cmidrule{2-6} & Random & -0.0637  & -0.0848  & -0.0832  & -0.0548  \\
        \cmidrule{2-6} & DQN & 0.0101  & -0.0092  & -0.0215  & -0.0504   \\ 
        \cmidrule{2-6} & SRC & \textbf{0.1288*}  & \textbf{-0.0042*}  & \textbf{-0.0159*}  & \textbf{0.2577*}  \\ 
    \bottomrule
    \end{tabular}
    }
\caption{Result on the industrial dataset.}
\label{table:Industrial}
\end{table}
\section{Conclusion}

In this paper, we formulate the path recommendation of the online education system as a set-to-sequence task and provide a new set-to-sequence ranking-based concept-aware framework, named SRC. Specifically, we first design a concept-aware encoder module that captures correlations between input learning concepts. The output is then fed to a decoder module that sequentially generates paths through an attention mechanism that handles correlations between learning and target concepts. The recommendation policy will be optimized through policy gradients. In addition, we introduce an auxiliary module based on knowledge tracking to enhance the stability of the model by evaluating students' learning effects on the learned concepts. We conduct extensive experiments on two real-world public datasets and one industrial proprietary  dataset, where SRC  demonstrates its performance superiority over other baselines. 

In future work, it might be an interesting direction to further explore the relationships between concepts, such as using graph neural networks. In addition, we plan to further deploy our model in the real-world online education system.

\section{Acknowledgements}
The SJTU team is partially supported by the National Natural Science Foundation of China (62177033). The work is also sponsored by Huawei Innovation Research Program.
We gratefully acknowledge the support of MindSpore \citep{mindspore}, CANN (Compute Architecture for Neural Networks), and Ascend AI Processor used for this research.
We also thank Li'ang Yin from Shanghai Jiao Tong University.

% \bibliographystyle{aaai23}
% \balance

\appendix


\begin{thebibliography}{52}
\providecommand{\natexlab}[1]{#1}

\bibitem[{Bian et~al.(2019)Bian, Wang, Liu, Lu, and Dong}]{bian2019adaptive}
Bian, C.-L.; Wang, D.-L.; Liu, S.-Y.; Lu, W.-G.; and Dong, J.-Y. 2019.
\newblock Adaptive learning path recommendation based on graph theory and an
  improved immune algorithm.
\newblock \emph{KSII Transactions on Internet and Information Systems (TIIS)},
  13(5): 2277--2298.

\bibitem[{Bruch et~al.(2020)Bruch, Han, Bendersky, and
  Najork}]{bruch2020stochastic}
Bruch, S.; Han, S.; Bendersky, M.; and Najork, M. 2020.
\newblock A stochastic treatment of learning to rank scoring functions.
\newblock In \emph{Proceedings of the 13th international conference on web
  search and data mining}, 61--69.

\bibitem[{Burges et~al.(2005)Burges, Shaked, Renshaw, Lazier, Deeds, Hamilton,
  and Hullender}]{burges2005learning}
Burges, C.; Shaked, T.; Renshaw, E.; Lazier, A.; Deeds, M.; Hamilton, N.; and
  Hullender, G. 2005.
\newblock Learning to rank using gradient descent.
\newblock In \emph{Proceedings of the 22nd international conference on Machine
  learning}, 89--96.

\bibitem[{Burges(2010)}]{burges2010ranknet}
Burges, C.~J. 2010.
\newblock From ranknet to lambdarank to lambdamart: An overview.
\newblock \emph{Learning}, 11(23-581): 81.

\bibitem[{Cai, Zhang, and Dai(2019)}]{cai2019learning}
Cai, D.; Zhang, Y.; and Dai, B. 2019.
\newblock Learning path recommendation based on knowledge tracing model and
  reinforcement learning.
\newblock In \emph{2019 IEEE 5th International Conference on Computer and
  Communications (ICCC)}, 1881--1885. IEEE.

\bibitem[{Cao et~al.(2007)Cao, Qin, Liu, Tsai, and Li}]{cao2007learning}
Cao, Z.; Qin, T.; Liu, T.-Y.; Tsai, M.-F.; and Li, H. 2007.
\newblock Learning to rank: from pairwise approach to listwise approach.
\newblock In \emph{Proceedings of the 24th international conference on Machine
  learning}, 129--136.

\bibitem[{Carbonell(1970)}]{carbonell1970ai}
Carbonell, J.~R. 1970.
\newblock AI in CAI: An artificial-intelligence approach to computer-assisted
  instruction.
\newblock \emph{IEEE transactions on man-machine systems}, 11(4): 190--202.

\bibitem[{Chang, Hsu, and Chen(2015)}]{chang2015modeling}
Chang, H.-S.; Hsu, H.-J.; and Chen, K.-T. 2015.
\newblock Modeling Exercise Relationships in E-Learning: A Unified Approach.
\newblock In \emph{EDM}, 532--535.

\bibitem[{Chen et~al.(2022)Chen, Huang, Tzeng, Lee, Huang, and
  Huang}]{chen2022personalized}
Chen, Y.-H.; Huang, N.-F.; Tzeng, J.-W.; Lee, C.-a.; Huang, Y.-X.; and Huang,
  H.-H. 2022.
\newblock A Personalized Learning Path Recommender System with Line Bot in
  MOOCs Based on LSTM.
\newblock In \emph{2022 11th International Conference on Educational and
  Information Technology (ICEIT)}, 40--45. IEEE.

\bibitem[{Cover and Hart(1967)}]{cover1967nearest}
Cover, T.; and Hart, P. 1967.
\newblock Nearest neighbor pattern classification.
\newblock \emph{IEEE transactions on information theory}, 13(1): 21--27.

\bibitem[{Deisenroth and Rasmussen(2011)}]{deisenroth2011pilco}
Deisenroth, M.; and Rasmussen, C.~E. 2011.
\newblock PILCO: A model-based and data-efficient approach to policy search.
\newblock In \emph{Proceedings of the 28th International Conference on machine
  learning (ICML-11)}, 465--472. Citeseer.

\bibitem[{Devlin et~al.(2018)Devlin, Chang, Lee, and
  Toutanova}]{devlin2018bert}
Devlin, J.; Chang, M.-W.; Lee, K.; and Toutanova, K. 2018.
\newblock Bert: Pre-training of deep bidirectional transformers for language
  understanding.
\newblock \emph{arXiv preprint arXiv:1810.04805}.

\bibitem[{Dwivedi, Kant, and Bharadwaj(2018)}]{dwivedi2018learning}
Dwivedi, P.; Kant, V.; and Bharadwaj, K.~K. 2018.
\newblock Learning path recommendation based on modified variable length
  genetic algorithm.
\newblock \emph{Education and information technologies}, 23(2): 819--836.

\bibitem[{Feng, Heffernan, and Koedinger(2009)}]{feng2009addressing}
Feng, M.; Heffernan, N.; and Koedinger, K. 2009.
\newblock Addressing the assessment challenge with an online system that tutors
  as it assesses.
\newblock \emph{User modeling and user-adapted interaction}, 19(3): 243--266.

\bibitem[{Friedman(2001)}]{friedman2001greedy}
Friedman, J.~H. 2001.
\newblock Greedy function approximation: a gradient boosting machine.
\newblock \emph{Annals of statistics}, 1189--1232.

\bibitem[{Grover et~al.(2019)Grover, Wang, Zweig, and
  Ermon}]{grover2019stochastic}
Grover, A.; Wang, E.; Zweig, A.; and Ermon, S. 2019.
\newblock Stochastic optimization of sorting networks via continuous
  relaxations.
\newblock \emph{arXiv preprint arXiv:1903.08850}.

\bibitem[{Hidasi et~al.(2015)Hidasi, Karatzoglou, Baltrunas, and
  Tikk}]{hidasi2015session}
Hidasi, B.; Karatzoglou, A.; Baltrunas, L.; and Tikk, D. 2015.
\newblock Session-based recommendations with recurrent neural networks.
\newblock \emph{arXiv preprint arXiv:1511.06939}.

\bibitem[{Hochreiter and Schmidhuber(1997)}]{hochreiter1997long}
Hochreiter, S.; and Schmidhuber, J. 1997.
\newblock Long short-term memory.
\newblock \emph{Neural computation}, 9(8): 1735--1780.

\bibitem[{Hu et~al.(2018)Hu, Da, Zeng, Yu, and Xu}]{hu2018reinforcement}
Hu, Y.; Da, Q.; Zeng, A.; Yu, Y.; and Xu, Y. 2018.
\newblock Reinforcement learning to rank in e-commerce search engine:
  Formalization, analysis, and application.
\newblock In \emph{Proceedings of the 24th ACM SIGKDD international conference
  on knowledge discovery \& data mining}, 368--377.

\bibitem[{Huang et~al.(2019)Huang, Liu, Zhai, Yin, Chen, Gao, and
  Hu}]{huang2019exploring}
Huang, Z.; Liu, Q.; Zhai, C.; Yin, Y.; Chen, E.; Gao, W.; and Hu, G. 2019.
\newblock Exploring multi-objective exercise recommendations in online
  education systems.
\newblock In \emph{Proceedings of the 28th ACM International Conference on
  Information and Knowledge Management}, 1261--1270.

\bibitem[{Inhelder et~al.(1976)Inhelder, Chipman, Zwingmann
  et~al.}]{inhelder1976piaget}
Inhelder, B.; Chipman, H.~H.; Zwingmann, C.; et~al. 1976.
\newblock \emph{Piaget and his school: a reader in developmental psychology}.
\newblock Springer.

\bibitem[{Joachims(2006)}]{joachims2006training}
Joachims, T. 2006.
\newblock Training linear SVMs in linear time.
\newblock In \emph{Proceedings of the 12th ACM SIGKDD international conference
  on Knowledge discovery and data mining}, 217--226.

\bibitem[{Joseph, Abraham, and Mani(2022)}]{joseph2022exploring}
Joseph, L.; Abraham, S.; and Mani, B.~P. 2022.
\newblock Exploring the Effectiveness of Learning Path Recommendation based on
  Felder-Silverman Learning Style Model: A Learning Analytics Intervention
  Approach.
\newblock \emph{Journal of Educational Computing Research}, 07356331211057816.

\bibitem[{Lee et~al.(2019)Lee, Lee, Kim, Kosiorek, Choi, and Teh}]{lee2019set}
Lee, J.; Lee, Y.; Kim, J.; Kosiorek, A.; Choi, S.; and Teh, Y.~W. 2019.
\newblock Set transformer: A framework for attention-based
  permutation-invariant neural networks.
\newblock In \emph{International conference on machine learning}, 3744--3753.
  PMLR.

\bibitem[{Li et~al.(2021)Li, Xu, Zhang, and Chang}]{li2021optimal}
Li, X.; Xu, H.; Zhang, J.; and Chang, H.-h. 2021.
\newblock Optimal hierarchical learning path design with reinforcement
  learning.
\newblock \emph{Applied psychological measurement}, 45(1): 54--70.

\bibitem[{Liu et~al.(2021)Liu, Wang, Ren, Wang, Yin, and Zhang}]{liu2021trap}
Liu, D.; Wang, S.; Ren, J.; Wang, K.; Yin, S.; and Zhang, Q. 2021.
\newblock Trap of Feature Diversity in the Learning of MLPs.
\newblock \emph{arXiv preprint arXiv:2112.00980}.

\bibitem[{Liu and Li(2020)}]{liu2020learning}
Liu, H.; and Li, X. 2020.
\newblock Learning path combination recommendation based on the learning
  networks.
\newblock \emph{Soft Computing}, 24(6): 4427--4439.

\bibitem[{Liu et~al.(2019)Liu, Tong, Liu, Zhao, Chen, Ma, and
  Wang}]{liu2019exploiting}
Liu, Q.; Tong, S.; Liu, C.; Zhao, H.; Chen, E.; Ma, H.; and Wang, S. 2019.
\newblock Exploiting cognitive structure for adaptive learning.
\newblock In \emph{Proceedings of the 25th ACM SIGKDD International Conference
  on Knowledge Discovery \& Data Mining}, 627--635.

\bibitem[{Long et~al.(2022)Long, Qin, Shen, Zhang, Xia, Tang, He, and
  Yu}]{long2022improving}
Long, T.; Qin, J.; Shen, J.; Zhang, W.; Xia, W.; Tang, R.; He, X.; and Yu, Y.
  2022.
\newblock Improving Knowledge Tracing with Collaborative Information.
\newblock In \emph{Proceedings of the Fifteenth ACM International Conference on
  Web Search and Data Mining}, 599--607.

\bibitem[{Luce(2012)}]{luce2012individual}
Luce, R.~D. 2012.
\newblock \emph{Individual choice behavior: A theoretical analysis}.
\newblock Courier Corporation.

\bibitem[{MindSpore(2022)}]{mindspore}
MindSpore. 2022.
\newblock MindSpore.
\newblock \url{https://www.mindspore.cn/}.

\bibitem[{Mnih et~al.(2013)Mnih, Kavukcuoglu, Silver, Graves, Antonoglou,
  Wierstra, and Riedmiller}]{mnih2013playing}
Mnih, V.; Kavukcuoglu, K.; Silver, D.; Graves, A.; Antonoglou, I.; Wierstra,
  D.; and Riedmiller, M. 2013.
\newblock Playing atari with deep reinforcement learning.
\newblock \emph{arXiv preprint arXiv:1312.5602}.

\bibitem[{Nabizadeh, Jorge, and Leal(2019)}]{nabizadeh2019estimating}
Nabizadeh, A.~H.; Jorge, A.~M.; and Leal, J.~P. 2019.
\newblock Estimating time and score uncertainty in generating successful
  learning paths under time constraints.
\newblock \emph{Expert Systems}, 36(2): e12351.

\bibitem[{Oosterhuis(2021)}]{oosterhuis2021computationally}
Oosterhuis, H. 2021.
\newblock Computationally efficient optimization of plackett-luce ranking
  models for relevance and fairness.
\newblock In \emph{Proceedings of the 44th International ACM SIGIR Conference
  on Research and Development in Information Retrieval}, 1023--1032.

\bibitem[{Pang et~al.(2020)Pang, Xu, Ai, Lan, Cheng, and Wen}]{pang2020setrank}
Pang, L.; Xu, J.; Ai, Q.; Lan, Y.; Cheng, X.; and Wen, J. 2020.
\newblock Setrank: Learning a permutation-invariant ranking model for
  information retrieval.
\newblock In \emph{Proceedings of the 43rd International ACM SIGIR Conference
  on Research and Development in Information Retrieval}, 499--508.

\bibitem[{Piaget and Duckworth(1970)}]{piaget1970genetic}
Piaget, J.; and Duckworth, E. 1970.
\newblock Genetic epistemology.
\newblock \emph{American Behavioral Scientist}, 13(3): 459--480.

\bibitem[{Piech et~al.(2015)Piech, Bassen, Huang, Ganguli, Sahami, Guibas, and
  Sohl-Dickstein}]{piech2015deep}
Piech, C.; Bassen, J.; Huang, J.; Ganguli, S.; Sahami, M.; Guibas, L.~J.; and
  Sohl-Dickstein, J. 2015.
\newblock Deep knowledge tracing.
\newblock \emph{Advances in neural information processing systems}, 28.

\bibitem[{Pinar et~al.(1995)Pinar, Reynolds, Slattery, Taubman
  et~al.}]{pinar1995understanding}
Pinar, W.~F.; Reynolds, W.~M.; Slattery, P.; Taubman, P.~M.; et~al. 1995.
\newblock \emph{Understanding curriculum: An introduction to the study of
  historical and contemporary curriculum discourses}, volume~17.
\newblock Peter lang.

\bibitem[{Plackett(1975)}]{plackett1975analysis}
Plackett, R.~L. 1975.
\newblock The analysis of permutations.
\newblock \emph{Journal of the Royal Statistical Society: Series C (Applied
  Statistics)}, 24(2): 193--202.

\bibitem[{Shao, Guo, and Pardos(2021)}]{shao2021degree}
Shao, E.; Guo, S.; and Pardos, Z.~A. 2021.
\newblock Degree planning with plan-bert: Multi-semester recommendation using
  future courses of interest.
\newblock In \emph{Proceedings of the AAAI Conference on Artificial
  Intelligence}, volume~35, 14920--14929.

\bibitem[{Shi et~al.(2020)Shi, Wang, Xing, and Xu}]{shi2020learning}
Shi, D.; Wang, T.; Xing, H.; and Xu, H. 2020.
\newblock A learning path recommendation model based on a multidimensional
  knowledge graph framework for e-learning.
\newblock \emph{Knowledge-Based Systems}, 195: 105618.

\bibitem[{Sun et~al.(2021)Sun, Zhuang, Zhu, He, and Xiong}]{sun2021cost}
Sun, Y.; Zhuang, F.; Zhu, H.; He, Q.; and Xiong, H. 2021.
\newblock Cost-effective and interpretable job skill recommendation with deep
  reinforcement learning.
\newblock In \emph{Proceedings of the Web Conference 2021}, 3827--3838.

\bibitem[{Swezey et~al.(2020)Swezey, Grover, Charron, and
  Ermon}]{swezey2020pirank}
Swezey, R.; Grover, A.; Charron, B.; and Ermon, S. 2020.
\newblock Pirank: Learning to rank via differentiable sorting.
\newblock \emph{arXiv preprint arXiv:2012.06731}.

\bibitem[{Tong et~al.(2020)Tong, Liu, Huang, Hunag, Chen, Liu, Ma, and
  Wang}]{tong2020structure}
Tong, S.; Liu, Q.; Huang, W.; Hunag, Z.; Chen, E.; Liu, C.; Ma, H.; and Wang,
  S. 2020.
\newblock Structure-based knowledge tracing: an influence propagation view.
\newblock In \emph{2020 IEEE International Conference on Data Mining (ICDM)},
  541--550. IEEE.

\bibitem[{Vinyals, Fortunato, and Jaitly(2015)}]{vinyals2015pointer}
Vinyals, O.; Fortunato, M.; and Jaitly, N. 2015.
\newblock Pointer networks.
\newblock \emph{Advances in neural information processing systems}, 28.

\bibitem[{Wang et~al.(2022)Wang, Liu, Wang, Wu, Fu, and Xie}]{wang2022multi}
Wang, X.; Liu, K.; Wang, D.; Wu, L.; Fu, Y.; and Xie, X. 2022.
\newblock Multi-level recommendation reasoning over knowledge graphs with
  reinforcement learning.
\newblock In \emph{Proceedings of the ACM Web Conference 2022}, 2098--2108.

\bibitem[{Williams(1992)}]{williams1992simple}
Williams, R.~J. 1992.
\newblock Simple statistical gradient-following algorithms for connectionist
  reinforcement learning.
\newblock \emph{Machine learning}, 8(3): 229--256.

\bibitem[{Xia et~al.(2008)Xia, Liu, Wang, Zhang, and Li}]{xia2008listwise}
Xia, F.; Liu, T.-Y.; Wang, J.; Zhang, W.; and Li, H. 2008.
\newblock Listwise approach to learning to rank: theory and algorithm.
\newblock In \emph{Proceedings of the 25th international conference on Machine
  learning}, 1192--1199.

\bibitem[{Yang et~al.(2020)Yang, Shen, Qu, Liu, Wang, Zhu, Zhang, and
  Yu}]{yang2020gikt}
Yang, Y.; Shen, J.; Qu, Y.; Liu, Y.; Wang, K.; Zhu, Y.; Zhang, W.; and Yu, Y.
  2020.
\newblock GIKT: a graph-based interaction model for knowledge tracing.
\newblock In \emph{Joint European Conference on Machine Learning and Knowledge
  Discovery in Databases}, 299--315. Springer.

\bibitem[{Zhang et~al.(2017)Zhang, Shi, King, and Yeung}]{zhang2017dynamic}
Zhang, J.; Shi, X.; King, I.; and Yeung, D.-Y. 2017.
\newblock Dynamic key-value memory networks for knowledge tracing.
\newblock In \emph{Proceedings of the 26th international conference on World
  Wide Web}, 765--774.

\bibitem[{Zhou et~al.(2018)Zhou, Huang, Hu, Zhu, and
  Tang}]{zhou2018personalized}
Zhou, Y.; Huang, C.; Hu, Q.; Zhu, J.; and Tang, Y. 2018.
\newblock Personalized learning full-path recommendation model based on LSTM
  neural networks.
\newblock \emph{Information Sciences}, 444: 135--152.

\bibitem[{Zhu et~al.(2018)Zhu, Tian, Wu, Shah, Chen, Ni, Zhang, Chao, and
  Zheng}]{zhu2018multi}
Zhu, H.; Tian, F.; Wu, K.; Shah, N.; Chen, Y.; Ni, Y.; Zhang, X.; Chao, K.-M.;
  and Zheng, Q. 2018.
\newblock A multi-constraint learning path recommendation algorithm based on
  knowledge map.
\newblock \emph{Knowledge-Based Systems}, 143: 102--114.

\end{thebibliography}
\end{document}